\documentclass[11pt,a4paper]{article}
\usepackage{graphicx}
\usepackage[latin1]{inputenc}
\usepackage{amsmath}
\usepackage{amsfonts}
\usepackage{amssymb}
\title{CAUSAL SPIN FOAMS AND THE ONTOLOGY OF SPACETIME}
\author{Miguel Lorente\thanks{E-mail adress: lorentemiguel@uniovi.es}\\Department of Physics, University of Oviedo\\
Avda. Calvo Sotelo 18, 33007 Oviedo, Spain
}
\date{Revised version: May 24, 2007}
\begin{document}
\maketitle 
\begin{abstract}
We present some modern theories on the structure of spacetime that can be classified as relational theories in the direction of Leibniz's ontology. In order to analyze the nature of spacetime we consider three levels of knowledge -observational, theoretical and ontological- to which the different models can be adscribed. Following similar approach to these models mentioned at the first sections, we present our theoretical model of the structure of spacetime, some physical applications, and the ontological interpretation of the model
\end{abstract}

\section{Introduction}

With surprise and satisfaction I received the invitation to participate in the Second International Conference on the Ontology of Spacetime, where explicit mention was made, not only to measurements and mathematical properties of some physical magnitude but to its metaphysical interpretations.

My approach to the nature of spacetime can be considered a relational theory, following the critical position of Leibniz (section 2) against the absolutist theory of Newton. In order to understand better my ontological interpretation I have presented some modern authors who have helped me to clarify the epistemological presuppositions as well as the ontological background that they have worked out. They are:
\begin{enumerate}
\renewcommand{\labelenumi}
{\roman{enumi}}
  \item[i)] the spin networks of Penrose (section 3),
  \item[ii)] the unified theory of elementary particles of D\"urr and Heisenberg (section 4),
  \item[iiii)] the spacetime code of Finkelstein (section 5),
  \item[iv)] the theory of simple alternatives (urs) of Weizs\"aker (section 6),
   \item[v)] the causal sets of Sorkin (section 7),
  \item[vi)] the quantum causal histories of Markopoulou (section 8),
  \item[vii)] the causal spin foams of Markopoulou and Smolin (section 9).
\end{enumerate}

In my opinion these models belong to the relational theories of spacetime, and their authors have tried the unification of quantum mechanics with the theory of relativity, in such a way that from the principles of Quantum Mechanics the structure of spacetime is derived, and the basis for the theory of relativity emerges.

For the presentation of our model we start with the epistemological presuppositions necessary to locate the concepts of space and time (section 10); it turns out that these are derived concepts. Then we present our model --the quantum causal lattice-- in two different examples, the cubic and the hyperbolic lattice (section 11). Because we want to make contact with physics we review some mathematical papers where we have translated the continuous language of physics into the discrete one (section 12).

Finally we describe the ontological background (underlying the theoretical level) made out of interactions of elementary beings, from which the concepts of spacetime are derived (section 13).

\section{Leibniz's relational theory of time and space}

In his correspondence with Clarke, Leibniz defended his position of a relational theory against Newton. Time is the order of those points not existing simultaneously and one is the ratio of the other. Space is the order of points that exist simultaneously and are connected by mutual interactions. Space is nothing more than the set of all points and their relations [1]. Position is that relation what the same is in different moments for different existing points and their coexisting relations with some particular points coincide completely [2]. A point changes its position if it changes its relations from some points to different ones. Motion is the change of different positions in time. Similar definitions of time and space have been given by Leibniz in his mathematical article ``Metaphysical principles of mathematics" [3], with the added concepts of duration and extension as magnitudes of time and space.

\subsection{The ideality of space}

The definitions given above have to be considered in some epistemological presuppositions.
As Earman says:

\newenvironment{margins}[2]
{
\begin{list}{} {
\setlength{\leftmargin}{#1}\begin{tiny}\label{} \end{tiny}
\setlength{\rightmargin}{#2}
} \item }
{\end{list}}

\begin{margins}{1.5cm}{0cm}
``There are passages from the 1680s in which Leibniz specifically refers to space and time as well-founded phenomena. Such passages only seem to compound the puzzle of the ideality thesis. 
The puzzle is resolved by noting that such passages disappear in the 1690s when Leibniz begins to make use of a trichotomy consisting of the monads, well-founded phenomena, and a third realm consisting of variously labeled `ideal',`neutral' and `imaginary'. It is to this third category that space and time are confined in Leibniz's later writings'' [4].
\end{margins}

According to this interpretation Leibniz's concepts of space does not correspond neither to the metaphysical level nor to the phenomenological level of observation and measurement, but to an intermediate level of knowledge that Leibniz calls ideal, and we call theoretical level.

\subsection{The monads}

In the ontological level Leibniz presents his Monadology. The monads correspond to the geometrical points in the theoretical (ideal) level. A monad is the metaphysical unity of a body (matter) with its entelechy (substantial form). Monads are the constituents of physical bodies, living organisms and human beings.

In the critical edition of the Monadology, Velarde explain the activities (action-passion) of monads:

\begin{margins}{1.5cm}{0cm}
``The internal principle is the force, as internal power of expansion, that generates a system (or state) of specific and internal qualities in each monad, called perception; and the action of internal principle that produces the change (or transit) from one to another state (from one to another perception) is called appetition (or appetite). Perception and appetition are thus metaphysical notions. Perception explains the monad as far as it is determinated from the world (passiv aspect), the appetition, by the contrary, presents it to us in its activ aspect, in the motion (tendency-impetus) from one to another perception. All the monads have a perceptive relation with the universe\ldots perception is nothing more but a plurality of relations of each monad with the rest"  [5] (traslation of M.L.)
\end{margins}

In the next sections we are going to present some models of contemporary physicists  who have followed the relational theory given by Leibniz about the nature of spacetime. We will present also the epistemological presuppositions underlying their models In their effort to justify their position they have reviewed the general principles that are involved in the unification of general relativity and quantum mechanics.

In order to achieve this goal it is unavoidable to understand better the nature of spacetime because it is a common ``arena"  for both theories.

\section{Penrose's spin networks}

We can consider Penrose's ideas of space and time very similar to the relational theory, because the concepts of space and time are derived from the set of interrelations among fundamental entities.

\subsection{Motivation}

Penrose considers the use of continuum in physics of mathematical utility rather than of essential physical necessity:

\begin{margins}{1.5cm}{0cm}
``I wish merely to point out the lack of firm foundation for assigning a physical reality to the conventional continuum concept" [6]
\end{margins}

The alternative proposed by Penrose is to derive the concepts of space and time from some combinatorial principles:

\begin{margins}{1.5cm}{0cm}
 ``My idea is to try to reformulate physical laws so that they may be expressed entirely in terms of quantities which are discrete according to quantum physics" [7].
\end{margins}

 In Penrose's model, quantum physics offers a collection of simple elements which are discrete, out of which the space and time can be derived.

\subsection{The Model}

For completness we summarize the main tracks of the theory of spin networks that can be founded in the primitive paper of Penrose [8].
The starting point is the total angular momentum of some fundamental unit (elementary particle or physical system). A set  of units are acting among themselves following the quantum rules of total angular momentum. In order to obtain a direction for some unit we need a large angular momentum such that the direction can be represented by the projection of the total spin $j$ on the m-component of the system in the direction of the z-axis. The value of the m-component of the unit can be obtained by combinatorial  process between two large units, that can be interpreted like the angle between the relative directions of both units.

The above picture is responsable for the emergency of eucliden geometry out of networks of units. If we want to implement this picture in the relativistic case, we have to introduce orbital components that require position of the units. Again, when two units are acting, a mutual displacement is emergent giving rise to the concept of position. So in the relativistic case a real 4-dimensional space-time is emergent from the interrelations of two or more elementary units.

\subsection{The background and the geometrical space}

Finally, the ontological status of the model is based in the interrelations between objects. As Penrose says:

\begin{margins}{1.5cm}{0cm}
``My model works with objects and the interrelations between objects. An object is thus ``locate" either directionally or positionally in terms of its relations with other objects. One does not really need a space to begin with. The notion of space comes out as a convenience at the end." [9]
\end{margins}

On this model Penrose makes a very important distinction between the auxiliary space (the background space) necessary to define total angular momentum in terms of spherical coordinates and the geometrical space that results from the angle between two spin directions. The last space is the real one and corresponds to the observed physical space. [10]

\section{Heisenberg's elementary length}

Although Heisenberg was one of the founders of Quantum Mechanics he never was satisfied with the orthodox interpretation and tried with some of his students and collaborators to modify the underlying ontology of Quantum principles. According to Weizs\"acker, who elaborated his dissertation with Heisenberg, the epistemological position of this is neither realistic (the realist  thinks he knows a priori what reality means) nor positivistic (the positivist thinks he knows a priori what experience means) and is closer to Kantian philosophy by which certain elements of the theory are taken to be preconditions of experience. In any sense, Heisenberg rejected the position taken by these three epistemologists because they presuppose the traditional opposition of observer and observed.

\begin{margins}{1.5cm}{0cm}
``When I was studying with him in Leipzig around 1930 he was already pondering on possible explanations of pure numbers like Sommerfeld's fine structure constants. According to this epistemological view there would have to be a new theory of elementary objects beyond general quantum theory\ldots His speculation of a fundamental length and even the introduction of the S-matrix belonged to this approach. In the meantime, reasons have become strong for the view that elementary particle theory is to be a perfect normal quantum theory which only limits the list of possible physical systems by additional axioms" [11] 
\end{margins}

\subsection{Heisenberg's unified theory of elementary particles}

The central assumption about additional axioms necessary to unify elementary particles from a quantum field theory is the invariance under some symmetry group. The model consists on a unique linear spinor field equation for a 4-component local Weyl-spinor-isospinor field operator which obeys non-canonical anticommutation relations. These spinors-isospinors fields are vector components of the representation of the Poincar\'e group and of the $U_1\times SU_2 $ group which (except for $U_1$) all occur in the generalized form of gauge symmetries of second kind. D\"urr has been working with Heisenberg in the detailed application of this model to the exuberant world of elementary particles.

The question that D\"urr put forward on Heisenberg Weltanschauung is whether or not there was another conceptually more basic level beneath our level of description that corresponds to the classification of elementary particles. 

\section{Finkelstein's space-time code}

D\"urr claims that Heisenberg was convinced of this hypothesis, and has tried to make connection of the more fundamental ideas of Weizs\"acker's urs or Finkelstein's space-time code with Heisenberg's unifying theory, in such a way that the continuous non linear field equation of Heisenberg is a limiting case of discrete models of Weizs\"acker or Finkelstein [12].

\subsection{Finkelstein's process theory}

According to Finkelstein the world is represented by a network of quantum processes which, in one version of his work [13], is built from "tetrads" as the only basic connected elements forming a structure with a checker board topology. The checker board constitutes the underlying structure of space-time manifold. This discrete structure can be considered the "arena" where the displacements and interactions of elementary particles take place. In particular D\"urr has shown that the non linear spinor Heisenberg's equation without its isopin degrees of freedom together with the hermitian conjugate equation (combined to a single non linear field equation for a 4-component hermitian Majorana spinor field) corresponds precisely to Finkelstein tetrad, in such a way that proceeding a discrete step in the checker board a new tetrad appears giving rise to the propagation in the network. [14]

Finkelstein has developed his program for the ontology of processes starting from the concept of monads, that reminds us of the Leibniz conception of material points out of which the structure of space and time is created. Finkelstein's ideas do not presuppose the existence of space and time, a similar position taken by Penrose [15].

\section{Weizs\"acker's ur hypothesis}

According to D\"urr, the unification of elementary particles proposed by Heisenberg was inspired in Weizs\"acker's theory of simple alternatives. This author, who tried first with Heisenberg a reconstruction of Quantum Mechanics, worked later with his group in Starnberg a new foundation of Quantum Mechanics. [16]

\subsection{Postulates for the Basic Structure of Quantum Mechanics}

Weizs\"acker new conception is derived from empirical and philosophical reflections on physical objects, that lead to some abstract, concrete and full Quantum Theories [17]. The abstract theory is constructed from Hilbert space, probability metric, rules of composition and dynamical laws. The concrete theory is constituted from real facts that can be reduced to simple alternatives, experiments yes/no, that are called urs, with the following properties: i) every experimental result can be reduced to a finite number of alternatives, ii) the number of possible alternatives is unlimited, iii) there are objects with only one alternative.

It corresponds to the full Quantum Theory to unify the abstract and concrete theory through a set of postulates, namely: i) Postulate of separable alternatives: there are alternatives whose states are separable from nearby all other states. Therefore we need only a finite number of alternatives to determine an object completely. ii) Postulate of indeterminacy: if a probability vector of the outcome of the possible alternative is defined, this must be a continuous function, in order that some state not belonging to the given alternative can be produced.

These postulates satisfy two conditions, which are accepted as preconditions of our experience: i) there is an actual infinity of future possibilities, and ii) no recourse to an actual infinity of facts is needed. In other words, the actual number of the simple alternatives in the universe is finite, but the number of possible alternatives in the future is infinite and they are governed by laws of probability. [18]

\subsection{The epistemology of Ur-hypothesis}

The goal of Weizs\"aker's reconstruction of quantum theory was to unify Quantum Mechanics with Theory of Relativity in such a way that the fundamental theory corresponds to the principles of Quantum Mechanics and the derived concepts correspond to the Theory of Relativity. In particular, if we accept the Ur-hypothesis as the fundamental one, the structure of space and time is a consequence of the former. The Hilbert space of the urs consists of the complex vector with two components (yes-no decisions). Its symmetry group is $SU(2)$, but this group is isomorphic to the group of rotations in the real three dimensional space, and this is the explanation why ordinary space is three-dimensional. [19]

The epistemological status of Ur-hypothesis can be understood if we locate it in the theoretical level between the empirical and the ontological one. In the theoretical level quantum physics and the Theory of Relativity are derived from the same principles. But can we adscribe a more fundamental level, an ontological level, to the Ur-hypothesis in such a way that there exist a correspondence between the former and the later as seen by a philosopher and a physicist? 

We recall the comment of G\"ornitz and Ischebeck on Weizs\"acker's Weltanschaung:

\begin{margins}{1.5cm}{0cm}
``The overwhelming success of science in the material world left no place for spirit in science. This process could be reversed by a quantum theory based on the foundations which Weizs\"acker has given, where matter and energy are united with information. On this basis it is possible that consciousness becomes a genuine part of natural science". [20]
\end{margins}

Similar approach to the philosophy of physics contained in the Ur-hypothesis is given by Lyre in his article on Weizs\"acker's Reconstruction of Physics. If one takes seriously Aristotle's Metaphysics all the substance in the world are composed out of matter (materia prima or hyle) and forms (forma substantialis or eidos). This last component can be identified with information and this is precisely what Weizs\"acker adscribes to the urs: a bit of information what it is obtained by the yes-no experiment. The Ur is reduced to the eidos in the ontological level. [21]

\section{Sorkin's causal sets}

The next two sections will deal with some variations of spin networks that take into account the causal relations among the elements of the discrete sets of events. In both cases, Sorkin and Markopoulou, introduced the hypothesis that there is a discrete reality underlying the continuous space. This reality or new substance is a causal set. Historically Sorkin claims that his ideas are rooted in Riemann's conception of discrete manifold, in which the principle of its metric relationships is already contained in the concept of the manifold itself. According to Sorkin, 

\begin{margins}{1.5cm}{0cm}
``The causal set is, of course, meant to be the deep structure of spacetime\ldots the spacetime cease to exist on sufficiently small scales and is superseded by an ordered discrete structure to which the continuum is only a coarse-grained, macroscopic approximation". [22]
\end{margins}

Sorkin's decision to accept causal sets was a reaction to the operationalism view of science, by which all the knowledge of nature are reduced to the set of operations by which we observe the experimental data. He accepted the ontological view that causal set is a real substratum, existing independently of any experimental activity of our part, and the elements of a causal set are real, and the notions of length and time emerge from relations among some fundamental entities. [23]

\subsection{A causal set and its embedding}

The discrete structure of space proposed by Riemann in 1854 was elaborated later by Robb [24] in 1914, where he proved that the geometry of 4-dimensional flat spacetime can be recovered from nothing more than the underlying points set and the order relation among points. Also Reichenbach [25] stressed the same fact and Finkelstein proposed in 1969 the original model of a causal set. [26]
As a mathematical structure, a causal set is a locally finite ordered set, i.e. a set $C$ endowed with a binary relation $<$ possessing the following three properties
i) Transitivity: $(\forall x,y,z \in C) (x<y<z \Rightarrow x<z)$
ii) Irreflexivity $(\forall x\in C) (x\nless x)$
iii) Local finiteness: $(\forall x,z\in C) (\mbox{card}\lbrace y\in C \vert x<y<z \rbrace < \infty)$. Property ii) implies the absence of cycles and property iii) is a formal way of saying that a causal set is discrete.

In order to compare the discrete causal set with the continuum space-time, Sorkin and collaborators introduce an embedding in such a way that i) causal relations among the points in the discrete are preserved in the continuous and ii) the embedded points are distributed uniformly with unit density. If these conditions are satisfied one can decompose the continuum manifold in elementary volumes such that to each one correspond one point, and in this way the Criterion of Riemann is fulfilled, that measuring is counting. [27]

\section{Markopoulou's quantum causal histories}

As we mention in the last section, Markopoulou makes the hypothesis that the underlying reality at the Planck's scale is discrete and it can be described by spin networks endowed with causal relations. This is appropriate to the canonical quantization of general relativity, in the sense of loop quantum gravity. Loop quantum gravity gives an exact microscopic description of spatial quantum geometry in term of basic states called spin networks. The dynamics is expressed in path integrals defined in terms of amplitudes for local moves along the spin networks. The basic operators of the theory (area and volume) are quantized. This construction, according to Markopoulou, suggests that at Planck scale geometry is discrete. Besides that, the theory is background independent, it does not live in a preexisting spacetime. As Smolin claims:

\begin{margins}{1.5cm}{0cm}
``At the end what is most satisfying about the picture of space given by loop quantum gravity is that it is completely relational. The spin networks do not live in space; their structure generates space. And they are nothing but a structure of relations, governed by how the edges are tied together at the nodes" [28]
\end{margins}

One of the main issues of loop quantum grvity is the problem of the low energy limit. From the fundamental combinatorial dynamics at low energy have to emerge the classical spacetime and the dynamics of general relativity. [29]

\subsection{Quantum causal histories}

Markopoulou has summarized several features in common with models of microscopic structure of spacetime. They are the following [30]

\begin{enumerate}
\renewcommand{\labelenumi}
{\roman{enumi}}
  \item[i)] At energies close to the Plank scale the Universe is discrete.
  \item[ii)]  Causality still persists; the Universe is described by the rules of causal sets presented by Sorkin et al. 
  \item[iii)]  Quantum theory is still valid at this level
  \item[iv)]  The model should be background independent
\end{enumerate}

These presuppositions are taken in account by Markopoulou to construct the quantum causal histories; for completeness we sketch them in the form of causal sets plus quantum operators [31]

\textbf{Causal set}: partially ordered set, locally finite, with preceding relation $\lbrace C, <\rbrace$     

Causal past: $ \lbrace r \mid r<p, r\in C \rbrace\equiv P(p)$

Causal future: $ \lbrace q \mid p<q, q\in C \rbrace\equiv  F(p)$ 

Set \textbf{a},  is a complete past of $p$ if every event in $P(p)$ is related to \textbf{a}.

Set \textbf{b},  is a complete future of $p$ if every event in $F(p)$ is related to \textbf{b}.

Two sets \textbf{a} y \textbf{b} are a complete pair if \textbf{a} is a complete past of \textbf{b} and \textbf{b} is a complete future of \textbf{a}.

\textbf{Quantum causal sets}: attach a Hilbert space to each event of a causal set representing elementary systems

\textbf{Quantum causal histories}: the evolution of a Quantum causal set is implemented by unitary operator between Hilbert spaces of a complete pair

\textbf{Quantum spin networks}: repeated applications of local moves takes one spin network into another.

Therefore, in quantum causal histories, with the Hilbert spaces on the events  and the operators on the causal relations, the quantum evolution strictly respects the underlying causal set. The ontological background of the spacetime --the quantum spacetime-- consists on a very large set of open systems joined by quantum operations, where unitary evolution arises only for a complete pair [32]

\section{Markopoulou and Smolin's causual spin foams}

Penrose's spin networks can be considered as a graph with edges labelled by irreducible representations of the group SU(2) and vertices labelled by intertwiners satisfying the rules of angular momenta. Similarly a spin foam is a 2-dimensional complex with faces labelled by irreducible representations of a group, generally the group SO(4) or SO(3,1), and edges labelled by intertwiners. (The 2-dimensional complex can be considered the 2-dimensional dual graph coming from the triangulation of a 4-dimensional manifold where to each 4-simplex corresponds a vertex and to each tetrahedron corresponds an edge). [33]

\subsection{Spin foam models}

With the help of spin foams one constructs spin foam models for quantum gravity that are intrinsecally discrete and are supposed to go in the low energy limit to the general relativity field equations and the continuous spacetime manifolds.

Several spin foam models have been proposed [34] such as Lorentian path integrals, string networks or topological quantum field theory. The most elaborated of them and thoroughly studied is the Barrett-Crane spin foam model. [35] 

\subsection{Causal spin foam models}

Using the kinematical setting of partition function for spin foams and the assumption of a micro-local causal structure (encoded in the orientation of spin netorks) Markopoulou and Smolin define a general class of causal spin foam model for quantum gravity [36]. The elementary transition amplitude for an initial spin network to another spin network is defined by a set of combinatorial rules.

Levine and Oriti have shown [37] that the Barrett- Crane model is the first non trivial example of a causal spin foam model, and that it represents a link between several areas or research, like canonical loop quantum gravity, sum over histories formulation, causal sets and dynamical triangulation.

\subsection{Background independent models}

Spin foam models have very important property. They are background independent quantum gravity models. They don't live in a pre-existing universe. They start with an underlying Planck scale quantum system with no reference to spatial temporal geometry. The geometry is defined intrinsically using subsystems and their relations. (Both quantum geometry and gravity emerges as a low energy continuous limit). In particular, spatial and temporal distances have to be defined internally by observers inside the system.

Markopoulou has developped a lengthy discussion on the topic [38] and has given several definitons, the first of which is: ``A theory is background independent if its basic quantities and concepts do not presuppose the existence of a given background spacetime metric".

This is consistent with the relational principle by which the metric has to be defined intenally, and in the case of a discrete manifold by counting the elements of the same manifold as Riemann claimed in his Inaugural Dissertation of 1854.

\section{Our model: epistemological presuppositions}

We can summarize the epistemological position of the authors mentioned in section 2 to 8 by three levels of human knowledge in the comprehension of the physical world. It will help to understand my own position in the interpretation of spacetime. [39]

\textbf{Level 1}: Physical magnitudes, such as distance, interval, mass, event, force, and so on, that are given by our sensations and perceptions.

\textbf{Level 2}: Theoreticall models, which are the generalization of metrical properties given by measurements and numerical relations among them.

\textbf{Level 3}: Fundamental concepts, representing the ontological properties of physical world given by our consciousness in an attempt to know the reality.

There must be some connections between the three levels. In Quantum Mechanics the theoretical models of microphysics in level 2 are related to observable magnitudes in level 1 by correspondence laws.

If we accept level 3 should be connected to level 2, an immediate questions is to ask about the justification of the rules governing the construction of theoretical systems. It would be ridiculous to postulate them as games rules. They must be grounded in properties of the world they want to describe. For instance the unification of Quantum Mechanics and the theory of Relativity should be made in level 2 where they belong to, but the underlying ontological concepts should be taken from level 3. 

We can now raise the following question: in level 1 we find primitive and derived concepts. According to philosophy of science it is almost impossible to decide whether some simple observale is primitive or not, because it depends on the type of experiment we have used to define it. Once we have decided the primitive concepts of a theory, the rest are derived concepts. The question to put forward: are the concepts of space and time primitive or derived concepts?

In absolute theories, space is a container where the particles are moving. Time is also a separated entity with respect to which the motion takes place. Therefore space and time are primitive concepts and can be thinked of in the absence of particles.

In relational theories, space and time consist on the set of relations of some fundamental objects. Obviously in this case, the concepts of space and time are derived. As Markopoulou explains: ``Spacetime geometry is a derivative concept and only applies in a approximate emergent level". [40]

This is a consequence of the relational character by which ``spatial and temporal distances are to be defined internally by observers inside the system".

 \section{A relational theory of spacetime: the causal cubic lattice}

Following the assumption of the last section now we give an explicit construction of a  formal
structure of spacetime, without the recourse to intuition. We can think of a set of
fundamental objects acting among themselves, giving rise to a  network of relations. These
relations do not pressupose some space. The objects are  nowhere if we consider them as
elements of the physical world in level 2. In order to  be specific we take as a naive network
a three-dimensional cubic lattice. Obviously the network can be taken with different
structure, such as, triangular,  quasiperiodic or random lattices. In order to make connection
with the euclidean geometry we take, for simplicity, a  infinite set of interacting points in
the relation 1 to 4, where one point is connected with no more and no less than four. 
The set of all relations form a  two-dimensional lattice, in which we can define:

A {\it path} is the connection between two different points, say, A and B, through 
points that are pairwise neighbours [41].

The {\it length} of a path is the numbers of points contained in the path, including the 
first and the last one.

A {\it minimal path} is a path with minimal length (in the picture the two paths 
between A and B are minimal). Between two point there can be different minimal 
paths.

\bigskip
\begin{center}
\includegraphics{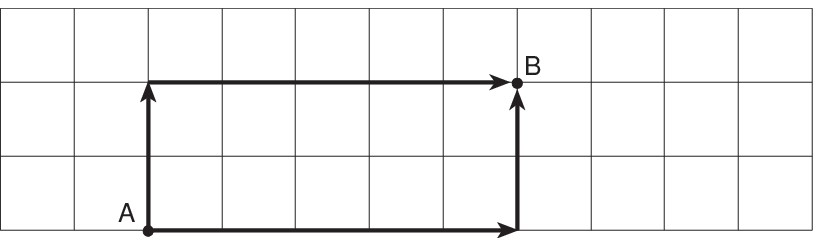}
\end{center}
\medskip

A {\it principal straight line} is a indefinite set of points in the lattice, such that each 
of them is contiguous to other two, and the minimal path between two arbitrary 
points of this line is always unique.

\vskip 13pt

{\noindent {\bf Theorem 1}. Through a point of a 2-dimensional square lattice pass only two 
different principal straight lines (they are called {\it orthogonal straight lines}).}

\vskip 13pt

{\noindent {\bf Theorem 2}. Two principal straight lines that are not orthogonal have all the
points  either in common or separated (in the last case they are called {\it paralell straight 
lines})}.

\vskip 13pt

From these two theorem we can define Cartesian (discrete) coordinates and an 
Euclidean space where the postulates of Hilbert can be applied (with the exception of 
the axioms of continuity). This structure of 2-dimensional space can be easily generalized to
3-dimensional cubic lattice. As we mentioned, those assumptions for the structure of space are
given in level 2, but  it corresponds to the properties of physical space described in level 1
by our  sensations.

In order to introduce the relation that correspond to time we  start with only two fundamental objects acting among themselves:

\bigskip
\begin{center}
\includegraphics{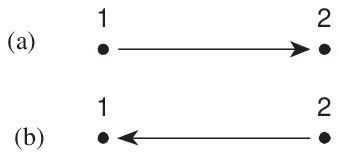}
\end{center}
\medskip

In (a), 1 is acting on 2, and in (b) 2 is acting on 1. But the action of 1 on 2 is supposed 
to be a necessary condition for the action of 2 on 1, and similarly the action of 2 on 1 
is supposed to be a necessary condition for a new action of 1 on 2. Thus we can think 
of a chain of mutual interactions arranged in a series of necessary conditions.
This picture has to be enlarged for the whole lattice. We take a set of interacting 
objects in the relations 1 to 2.

\bigskip
\begin{center}
\includegraphics{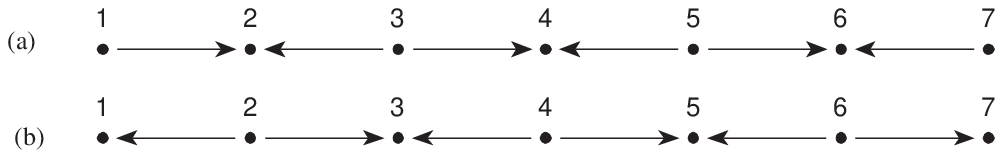}
\end{center}
\medskip

In (a), 1 is acting on 2, 3 is acting on 2 and 4, 5 is acting on 4 and 6, 7 is acting on 6.
In (b), 2 is acting or 1 and 3, 4 is acting on 3 and 5, 6 is acting on 5 an 7.

We postulate that the actions of (a) are necessary conditions for the actions of (b) and 
the actions of (b) are necessary conditions for a further action of type (a) an so on.

Now take a network of objects acting in the relation 1 to 4.

\bigskip
\begin{center}
\includegraphics{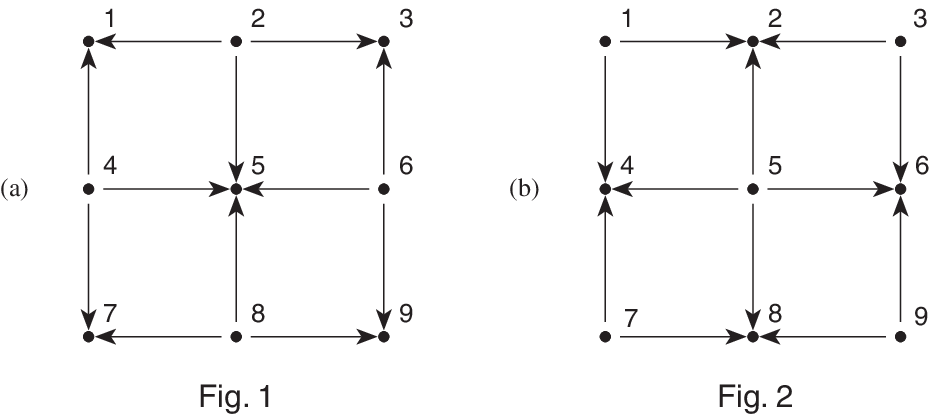}
\end{center}
\medskip

In (a), 2 is acting on 1, 3, 5; 4 is acting on 1, 5, 7; 6 is acting on 3, 5, 9; 8 is acting on 
5,7,9. In (b), 1 is acting on 2 and 4; 3 is acting on 2 and 6; 5 is acting on 2, 4, 6, 8; 7 is
acting  on 4 and 8; 9 is acting on 6 and 8. As before we postulate that the actions of (a) be
necessary conditions for the actions  of (b) and so on. These logical properties of
interactions belong to level 2 and do not pressupose the  concept of time, but they can be put
in correspondence with the physical properties of  time given in level 1.

Similar causal relations can be assumed in the hyperbolic lattice. [42]

\section{Physical and philosophical implications of the model}

The assumption of our model of spacetime implies some physical consequences for the classical as well for the quantum physics:

\begin{itemize}
 \item[i)] The space time is discrete, therefore the physical laws are written in the language of finite differences. The solutions have to be described by continuous functions of discrete variable. We present some particualr example in [43]
 \item[ii)] The symmetry of the model, in case of a Minkowski hypercubic lattice, is still Poincar\'e group, although one has to select those integral transformations that keep the lattice invariant. [44]
 \item[iii)] Lattice gauge theories are not only a mathematical tool but a realistic theory, because they correspond to the underlying discrete structure of spacetime. [45]
 \item[iv)] In General Relativity Riemannian manifolds have to be substituted by discrete graphs where geometrical magnitudes like metric tensor, curvature, have to be calculated by intrinsic properties of the graph. [46]
 \item[v)] In order to study the topology of a graph we embed it on a continuous manifold. Some quantum gravity models are based on this technique: the underlying spacetime is discrete, but its embedding is continuous, where field representations are attached. [47]
\end{itemize}

In our model from the data of our observations we have constructed theoretical models with the help of which we can give explanations and make predictions.

We have substituted the physical structure of spacetime by some network of interrelations among fundamental entities fom which the concept of spacetime emerges.

In order to deepen in the nature of spacetime we suppose there is an ontological level (level 3) where the physical properties of level 2 are interpreted with the metaphysical principles of the material objects.

In other models we have reviewed in the first sections we find some epistemological presuppositions to better understand their model. In Penrose's conception there are three ``worlds" (inspired in Popper's philosophy) that correspond to the physical, the mental, and Platonic mathematical objects, such that all of them are cyclically and misteriously connected [48]. The difference with our epistemological scheme is that the Platonic or mathematical world which would correspond to our ontological level is lying in an ideal world outside of the physical one.

In the causal set model Sorkin presupposes that a physical theory passes through three stages: an initial stage in which a particular ``substance" or type of matter presents itself in a characteristic group of phenomena; a second stage in which the new substance is clearly discerned in relation to the phenomena; and a final stage in which the comprehensive dynamics characterizing this substance is understood. [49] Sorkin claims that the new substance underlying space-time is a causal set, and he is convinced, contrary to operationalist ontology, that the elements of causal sets are real. [50] Therefore in Sorkin's model we find three level of knowledge: phenomena, physical theory and reality.

In the causal spin foam model and in the quantum causal histories there is a combination of causal sets with quantum mechanics in such a way that in the first model the non existence of the wave function of the universe imposes causality to the quantum subsistems, and in the second one the Hilbert spaces are attached to the events of the causal set. We find these two models very similar to our model with respect to the first and second level but they lack some ontological interpretation that we are going to present in the next section for our model.

\section{An Ontological interpretation of the nature of spacetime}

	In our epistemological presuppositions for the interpretation of spacetime, we have postulated the level 3 as the ontological background of the theoretical models of level 2. In a relational theory of the nature of spacetime, the concept of substance should be adscribed to the fundamental entities -monads, urs, units, events- the interaction of which give rise to the set of relations responsible of the structure of spacetime.

	Is it possible to make some Ansatz about the nature of these fundamental objects? If we take the extension as the first property of matter, as Descartes has claimed, space and time should be considered necessary at the beginning of a fundamental theory.. We prefer the point of view that the most essential property of material objects is the capacity of producing effects in other objects, which was identified by Leibniz with the concept of force. [51]

	There is a causal relation between the force and the effect (the principle of external causality in Aristotelian philosophy). The set of all causal relations among the fundamental objects can be taken as the ontological background in level 3 for the relational theories of spacetime, such as Penrose's spin networks, Sorkin's causal sets, Markopoulou's causal quantum histories.

	But still the picture is not complete. When the principle of causality is applied in classical mechanics or in the theory of special or general relativity it is supposed to follow the law of determinism: Given some mechanical system under particular initial conditions the same forces will produce always the same effects. 

If we want to implement the principle of causality with quantum effects, as in the causal spin foams, we have to introduce the probability laws in the production cause-effect, as required by the postulates of quantum mechanics. Coming back to the level 3 the ontology of material objects is characterized not only by the principle of causality but also by the laws of probability. [52]

\subsection{Analogies and differences}

We have presented the ontological status of our model and now we want to compair it with others we have mentioned before. All of them start from some fundamental entities --monads, processes, units, urs, events-- the interrelations of which produce a network responsable for the emergency of spacetime.

The set of these elementary entities is locally finite, a condition necessary for the discretness of spacetime. It means that each elementary entity is individually separated from the rest. [53]

Causality is a fundamental property for all entities and is responsable of the interactions among them. All the causal spacetime networks can be reduced to the evolution (in discrete time) of several causal networks for different discrete time values. [54] In particular the evolution in time of the causal cubic lattice (section 11) can be reduced to two causal sets (figures 1 and 2) for two different discrete times.

But we have also detected some discrepancies. First of all in their models there is an identification of epistemological level 2 and 3, or, even more, they don't mention the ontological level because the elementary entities are reduced to some physical effects --simple alternatives, yes/no experiments, combinations of two angular mementa, events-- where no mention of the ontological status is given.

Secondly, the lack of substantive character of elementary entities makes very difficult to predict the situation of these entities after the causal effect has taken place. Do they disappear? Are they transformed in other entities? In our model the action of some elementary being produces some effect in other being, but both beings persist in their existence. As a consequence the underlying network elementary beings persist in time.

A third difference is concerned with the embedding. In other models causal sets are embedded in some continuous manifold, such that we can talk about some elementary length between two different events connected by some causal action. In our model there is no elementary length because the distance between two causal interactions is reduced to the process of counting. Therefore a causal set of interactions among elementary beings can be embedded in an instant of time and in a geometrical point.

\section{Concluding remarkes}

	We have presented a model on the nature of spacetime that turns out to be relational, discrete, causal and quantum, the ontological background of which consists on the causal interactions of material individual beings, that we call ``hylions" [55]

Our model can be considered, from the physical point of view, a particular example of the causal spin foam models, that we have explained in detail, although there are some ontological differences between  both of them.

The ontological interpretations of our model should be deepened in two directions: the nature of material beings and the relations to other beings in the Universe. To do this work is encouraging us the Spacetime Society.

\subsection*{Acknowledgements}

	This work has been partially supported by Ministerio de Educación y Ciencia, grant BFM2003-00313/FIS


\begin{thebibliography}{99}

\bibitem{} Leibniz, ``Correspondence with Samuel Clarke",  III letter, n.4.

\bibitem{}  Leibniz, ``Correspondence with Samuel Clarke", V letter, n. 47.

\bibitem{} G.W. Leibniz, ``Initia rerum mathematicarum metaphysica" in Mathematische Schriften (C. I. Gerhardt, ed.) Olms, Hildesheim 1962, vol. VII, p. 17-39.

\bibitem{} Earman, \textit{World enough and space-time} MIT Press, Cambridge 1989, p. 15.

\bibitem{} Monadolog\'ia, Edici\'on de J. Velarde, Biblioteca Nueva, Madrid 2001, p. 47.

\bibitem{} R. Penrose, ``On the nature of quantum geometry" in \textit{Magic without magic} (J.A. Wheeler, J.R. Klauder, ed.) Freeman, San Francisco 1972, p. 334

\bibitem{} See reference [6], p 335

\bibitem{} R. Penrose, ``Angular momentum: an approach to combinatorial space-time", in \textit{Quantum theory and beyond} (T. Bastin, ed.) Cambridge U. Press, Cambridge 1971

\bibitem{} Reference [8], p. 174

\bibitem{} Reference [6], p. 339

\bibitem{} C. F. von Weizs\"acker, ``Heisenberg's conception of Physics", in \textit{Quantum Theory and the Structure of Spacetime}, (L. Castell et al. ed.) Hanser, Munich, vol. 2, p. 18

\bibitem{} H. P. D\"urr, ``Heisenberg's unified theory of elementary particles and the structure of time and space", \textit{Quantum theory and the structure of spacetime} (L. Castell, M. Drieschner, C. F. Weizs\"acker, ed.) Hanser, M\"unchen 1977, vol. 2, p. 33

\bibitem{} D. Finkelstein, ``Space-time code" \textit{Phys. Rev. 184}, 1261-1271 (1969)

\bibitem{} H. P. D\"urr, Reference [12] p. 40

\bibitem{} D. Finkelstein ,``Space-time code IV" \textit{Phys. Rev. D9}, 2219 (1974)

\bibitem{} L. Castell, M. Drieschner, C. F. Weizs\"aker (eds.), \textit{Quantum theory and the structure of Space and Time} (6 volumes) Hanser, Munich 1975-1986

\bibitem{} Th. G\"ornitz, O. Ischebeck, ``An introduction to C.F. von Weizs\"aker's Program for a reconstruction of Quantum Theory", in \textit{Time, Quantum and Information} (L. Castell, O. Ischebeck ed.), Springer, Berlin, 2003

\bibitem{} C. F. v. Weizs\"acker, ``A Reconstruction of Quantum Theory", in \textit{Quantum Theory and the structure of time and space} (L. Castell et al. ed.) Hanser, M\"unchen, vol. 3, p. 7

\bibitem{} C. F. v. Weizs\"acker, ``Binary alternatives and the space-time structure" in \textit{Quantum Theory and the structure of time and space} (L. Castell, et al. ed.) Hanser, M\"unchen 1977, vol. 2, p. 86

\bibitem{} See Reference [17], p. 268

\bibitem{} H. Lyre, ``C. F. von Weizs\"acker's Reconstruction of Physics: Yesterday, Today, Tomorrow" in \textit{Time, Quantum and Information} (L. Castell, O. Ischebeck ed.) Springer, Berlin 2003, p. 381

\bibitem{} R. Sorkin, ``Causal sets: Discrete Gravity" (Notes for the Valdivia Summer School Jan. 2002) arXiv: gr-qc/0309009

\bibitem{} R. Sorkin, ``A specimen of theory construction from quantum gravity" in \textit{The Creation of Ideas in Physics} (Jarrett Leplin ed.) Kluwer, Dordrecht 1995. arXiv: gr-qc/9511063

\bibitem{} A. Robb, ``A theory of Time and Space", Cambridge U. Press 1914

\bibitem{} H. Reichenbach, ``Axiomatization of the theory of relativity", University of California Press, Berkeley 1969

\bibitem{} See Reference [13]

\bibitem{} L. Bombelli, J. Lee, D. Meyer, R. Sorkin, ``Space-Time as a causal set" \textit{Phys. Rev. Lett 59}, 521-524 (1987)

\bibitem{} L. Smolin, \textit{Three roads to quantum gravity}, Weidenfeld and Nicholson, London 2000, p. 138

\bibitem{} F. Markopoulou, L. Smolin, ``Disordered locality in loop quantum gravity states", arXiv:gr-qc/0702044

\bibitem{} F. Markopoulou, ``Planck-scale models of the universe" \textit{Science and ultimate reality: quantum theory, cosmology and complexity} (J. D. Barrow, P. Davies, C. Harper ed.) C.U.P., Cambridge 2003, arXiv:gr-qc/0210086

\bibitem{} F. Markopoulou, ``Quantum causal histories", \textit{Class, Quant. Grav. 17}, 2059-2072 (2000) arXiv:hep-th/9904009

\bibitem{} F. Markopoulou, ``And insider's guide to quantum causal histories", arXiv: hep-th/9912137

\bibitem{} J. Baez, ``An introduction to Spin Foam Models of Quantum Gravity and BF Theory", \textit{Lect. Notes Phys. 543} 25-94 (2000)

\bibitem{} A. P\'erez, ``Spin Foam Models for Quantum Gravity", \textit{Class. Quant. Grav. 20} (2003) R43 arXiv: gr-qc/0301113

\bibitem{} J. W. Barrett and L. Crane, ``Relativistic Spin Networks and Quantum Gravity", \textit{J. Math. Phys. 39} (1998), 3296-3302

\bibitem{} F. Markopoulou and L. Smolin, ``Quantum Geometry with Intrinsic Local Causality", \textit{Phys. Rev. D 58}  084032 (1998) ; ``Causal Evolution of Spin Networks", \textit{Nucl. Phys. B 508} (1997), 409-430

\bibitem{} E. R. Livine, D. Oriti, ``Implementing causality in the spin foam quantum geometry", \textit{Nucl. Phys. B } (2003), 231-279, arXiv:gr-qc/0210064

\bibitem{} F. Markopoulou,``New directions in Background Independent Quantum Gravity",  arXiv: gr-qc/0703097

\bibitem{} M. Lorente, ``A realistic interpretation of Lattice gauge theories'' in \textit{Fundamental Problems in quantum physics} (M. Ferrero, van der Merwe, ed.) Kluwer, New York 1995, p. 177-186, arXiv: hep-lat/0312044

\bibitem{} See Reference [38]

\bibitem{} M. Lorente, ``Quantum process and the foundation of relational theories of spacetime'', in \textit{Relativity in general} (J. D\'iaz Alonso, M. Lorente, ed.) Paris: Editions Frontieres 1994, p. 297-302, arXiv: gr-qc/0312119 

\bibitem{} M. Lorente, ``A discrete curvature on a planar graph"  \textit{Encuentros de F\'isica Fundamental Alberto Galindo} U.C.M. Madrid 2004, p. 339-353, arXiv: gr-qc/0412094

\bibitem{} M. Lorente, ``Continuous vs. discrete models for the quantum harmonic oscillator and the hydrogen atom'' Phys. Lett. A 285(2001) 119-126, arXiv: quant-ph/0401087

\bibitem{} M. Lorente, P. Kramer, ``Representations of the discrete inhomogeneous Lorentz group and Dirac wave equation on the lattice'', \textit{J. Phys A: Mat. Gen. 32} (1999) 2481-2497, arXiv: hep-lat/0401019

\bibitem{} M. Lorente, ``A new scheme for the Klein-Gordon and Dirac fields on the lattice with axial anomaly'', \textit{J. Group. Theor. Phys, 1} (1993) 105-121, arXiv: hep-lat/0312039

\bibitem{} L. Bombelli, M. Lorente. ``A combinatorial approach to discrete geometry'' Proceed XXVIII Encuentros Relativistas Espa\~noles (J. D\'i­az, L. Mornas, ed.) Oviedo 2005, arXiv: gr-qc/0512142

\bibitem{} P. Kramer, M. Lorente, ``Surface embedding, topology and dualization for spin networks'', \textit{J. Phys A: Mat. Gen. 35} (2002) 8563-8574

\bibitem{} R. Penrose, \textit{The road to reality}, Knopf. N.Y. 2005, p. 17

\bibitem{} See Reference [27]

\bibitem{} See Reference [23]

\bibitem{} Ref. [5] p. 23

\bibitem{} M. Lorente, ``A causal interpretation of the structure of space and time'' in \textit{Foundations of Physics} (P. Weingartner and G. Dorn, ed.) Vienna: Verlag H\"{o}lder-Pichler-Tempsky 1986. P.345-368.

\bibitem{} See Reference [27] p. 522

\bibitem{} F. Markopoulou, L. Smolin, ``Causal evolution of spin networks", \textit{Nucl. Phys. B 508} (1997) p. 409-430

\bibitem{} Ref. [52] p. 362



\end{thebibliography}
\end{document}